\documentclass{aa}
\usepackage[T1]{fontenc}
\usepackage{xcolor}
\usepackage{url}
\usepackage{amsmath}
\usepackage{amssymb}
\usepackage{graphicx}
\PassOptionsToPackage{normalem}{ulem}
\usepackage{ulem}

\makeatletter

\newcommand{\lyxdot}{.}

\usepackage{txfonts}  
\bibpunct{(}{)}{;}{a}{}{,}  
\usepackage[dvipdfm, breaklinks, colorlinks, citecolor=blue]{hyperref} 

\usepackage{color}

\usepackage{listings}
\usepackage[scaled=0.82]{beramono}  

\definecolor{deepblue}{rgb}{0,0,0.5}
\definecolor{deepred}{rgb}{0.6,0,0}
\definecolor{deepgreen}{rgb}{0,0.5,0} 

\renewcommand{\imath}{\mathrm{i}} 
\DeclareMathOperator{\floor}{floor}
\DeclareMathOperator{\sign}{sign}

\makeatother

\begin{document}

\title{CORDIC-like method for solving Kepler's equation}

\abstract{Many algorithms to solve Kepler's equations require the evaluation
of trigonometric or root functions.}{We present an algorithm to
compute the eccentric anomaly and even its cosine and sine terms without
usage of other transcendental functions at run-time. With slight modifications
it is also applicable for the hyperbolic case.}{Based on the idea
of CORDIC, our method requires only additions and multiplications and a short
table. The table is independent of eccentricity and can be hardcoded.
Its length depends on the desired precision.}{The code is short.
The convergence is linear for all mean anomalies and eccentricities
$e$ (including $e=1$). As a stand-alone algorithm, single and double
precision is obtained with 29 and 55 iterations, respectively. Half or two-thirds of the iterations can be saved in combination with
Newton's or Halley's method at the cost of one division.}{}

\author{M.~Zechmeister\inst{1}}

\institute{Institut f\"ur Astrophysik, Georg-August-Universit\"at, Friedrich-Hund-Platz
1, 37077 G\"ottingen, Germany\\
\email{zechmeister@astro.physik.uni-goettingen.de}}

\date{Received 4 April 2018 / Accepted 14 August 2018}
\maketitle

\section{Introduction}

\begin{figure}
\begin{centering}
\includegraphics[width=1\linewidth]{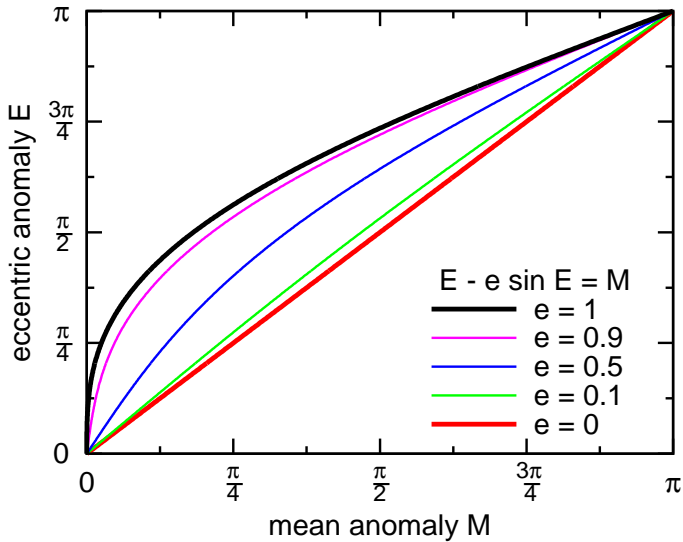}
\par\end{centering}
\caption{\label{fig:KE}Kepler's equation for five different eccentricities.}
\end{figure}

Kepler's equation relates the mean anomaly $M$ and the eccentric
anomaly $E$ in orbits with eccentricity $e$. For elliptic orbits
it is given by
\begin{equation}
E-e\sin E=M(E).\label{eq:KE}
\end{equation}
The function is illustrated for eccentricities 0, 0.1, 0.5, 0.9, and~1
in Fig.~\ref{fig:KE}. It is straightforward to compute $M(E)$.
But in practice $M$ is usually given and the inverse function $E(M)$
must be solved.

The innumerable publications about the solution of Kepler's equation
\citep{1993sket.book.....C} highlight its importance in many fields
of astrophysics (e.g. exoplanet search, planet formation, and star cluster
evolution), astrodynamics, and trajectory optimisation. N-body hybrid
algorithms also make use of this analytic solution of the two-body
problem \citep{1991AJ....102.1528W}. Nowadays, computers can solve
Eq.~(\ref{eq:KE}) quickly. But the speed increase is counterbalanced
by large data sets, simulations, and extensive data analysis (e.g.
with Markov chain Monte Carlo). This explains ongoing efforts to accelerate
the computation with software and hardware, for example by parallelising
and use of graphic processing units (GPUs; \citealp{2009NewA...14..406F}).

The Newton-Raphson iteration is a common method to solve Eq.~(\ref{eq:KE})
and employs the derivative $E'$. In each step $n$ the solution is
refined by
\begin{equation}
E_{n+1}=E_{n}+E'(M_{n})(M-M_{n})=E_{n}-\frac{E_{n}-e\sin E_{n}-M}{1-e\cos E_{n}}.\label{eq:Newton}
\end{equation}
A simple starting estimate might be $E_{0}=M+0.85e$ \citep{1987CeMec..40..303D}.
Each iteration comes at the cost of evaluating one cosine and one
sine function. Hence one tries to minimise the number of iterations.
This can be done with a better starting estimate. For example, \citet{1995CeMDA..63..101M}
provides a starting estimate better than $10^{-4}$ by inversion of a
cubic polynomial which however requires also transcendental functions
(four roots). \citet{BOYD200712} further polynomialise Kepler\textquoteright s
equation through Chebyshev polynomial expansion of the sine term and
yielded with root finding methods a maximum error of $10^{-10}$ after
inversion of a fifteen-degree polynomial. Another possibility to
reduce the iteration is to use higher order corrections. For instance
Pad\'{e} approximation of order {[}1/1{]} leads to Halley's method.

Pre-computed tables can be an alternative way for fast computation
of $E$. \citet{1997CeMDA..66..309F} used a table equally spaced
in $E$ to accelerate a discretised Newton method, while \citet{2006CeMDA..96...49F}
proposed an equal spacing in $M$, that is, a direct lookup table which
must be $e$-dependent and therefore two-dimensional. Both tables
can become very large depending on the desired accuracy.

So the solution of the transcendental Eq.~(\ref{eq:KE}) often comes
back to other transcendental equations which themselves need to be
solved in each iteration. This poses questions as to how those often
built-in functions are solved and whether there is a way to apply
a similar, more direct algorithm to Kepler's equation. 

The implementation details of those built-in functions are hardware
and software dependent. But sine and cosine are often computed with
Taylor expansion. After range reduction and folding into an interval
around zero, Taylor expansion is here quite efficient yielding $10^{-16}$
with 17th degree at $\frac{\pi}{4}$. Kepler's equation can also
be Taylor expanded \citep{1968NASTN4460.....S} but that is less efficient,
in particular for $e=1$ and around $M=0$ where the derivative becomes
infinite. Similarly, root or arcsine functions are other examples
where the convergence of the Taylor expansion is slow. For root-like
functions, one again applies Newton-Raphson or bisection methods.
The arcsine can be computed as $\arcsin(x)=\arctan(\frac{x}{\sqrt{1-x^{2}}})$
where Taylor expansion of the arctangent function is efficient and
one root evaluation is needed.

An interesting alternative method to compute trigonometric functions
is the Coordinate Rotation Digital Computer (CORDIC) algorithm which
was developed by \citet{volder1959cordic} for real-time computation
of sine and cosine functions and which for instance found application
in pocket calculators. CORDIC can compute those trigonometric and
other elementary functions in a simple and efficient way.
In this work we study whether and how the CORDIC algorithm can
be applied to Kepler's equation.

\section{\label{sec:CORDIC-KE}Scale-free CORDIC algorithm for Kepler's
equation}

Analogous to CORDIC, we compose our angle of interest, $E$, by a
number of positive or negative rotations $\sigma_{i}=\pm1$ with angle
$\alpha_{i}$
\begin{equation}
E_{n}=\sum_{i=1}^{n}\sigma_{i}\alpha_{i}.\label{eq:En}
\end{equation}
The next iteration $n$ is then
\begin{equation}
E_{n+1}=E_{n}+\sigma_{n+1}\alpha_{n+1}.\label{eq:En+1}
\end{equation}
The rotation angle $\alpha_{n}$ is halved in each iteration
\begin{equation}
\alpha_{n+1}=\frac{\alpha_{n}}{2}=\frac{\alpha_{1}}{2^{n}}\label{eq:alpha}
\end{equation}
so that we successively approach $E$. The direction of the next rotation
$\sigma_{n+1}$ is selected depending on the mean anomaly $M_{n}$
calculated from $E_{n}$ via Eq.~(\ref{eq:KE}). If $M_{n}=E_{n}-e\sin E_{n}$
overshoots the target value $M$, then $E_{n+1}$ will be decreased,
otherwise increased 
\begin{equation}
\sigma_{n+1}=\begin{cases}
-1 & E_{n}-e\sin E_{n}>M\\
+1 & \text{else}.
\end{cases}\label{eq:sgn}
\end{equation}

Equations (\ref{eq:En+1})\textendash (\ref{eq:sgn}) provide the iteration
scheme which so far is a binary search or bisection method (with respect
to $E$) to find the inverse of $M(E)$. Choosing as start value $E_{0}=0$
and start rotation $\alpha_{1}=\frac{\pi}{2}$, we can cover the range
$E\in[-\pi,\pi]$ and pre-compute the values of rotation angles with
Eq.~(\ref{eq:alpha})
\begin{align}
\alpha_{n} & =\frac{\pi}{2^{n}}.\label{eq:alpha_n}
\end{align}

The main difference of our algorithm from the CORDIC sine is the modified decision
for the rotation direction in Eq.~(\ref{eq:sgn}) consisting of the additional term $e\sin E$. For the special case $e=0$, the eccentric and mean
anomalies unify and the iteration converges to $E_{n}\rightarrow E=M$.

One major key point of CORDIC is that, besides the angle $E_{n}$,
it propagates simultaneously the Cartesian representation, which are
cosine and sine terms of $E_{n}$. The next iterations are obtained
through trigonometric addition theorems 
\begin{align}
\cos E_{n+1} & =\cos(E_{n}+\sigma_{n+1}\alpha_{n+1})\label{eq:rotE-1}\\
 & =\cos E_{n}\cos\alpha_{n+1}-\sigma_{n+1}\sin E_{n}\sin\alpha_{n+1}\\
\sin E_{n+1} & =\sin(E_{n}+\sigma_{n+1}\alpha_{n+1})\\
 & =\sigma_{n+1}\cos E_{n}\sin\alpha_{n+1}+\sin E_{n}\cos\alpha_{n+1},
\end{align}
which is a multiplication with a rotation matrix
\begin{equation}
\left(\begin{array}{c}
c_{n+1}\\
s_{n+1}
\end{array}\right)=\left[\begin{array}{cc}
\cos\alpha_{n+1} & -\sigma_{n+1}\sin\alpha_{n+1}\\
\sigma_{n+1}\sin\alpha_{n+1} & \cos\alpha_{n+1}
\end{array}\right]\left(\begin{array}{c}
c_{n}\\
s_{n}
\end{array}\right).\label{eq:rotE}
\end{equation}
Here we introduced the abbreviations $c_{n}=\cos E_{n}$ and $s_{n}=\sin E_{n}$.

\begin{figure}
\begin{centering}
\includegraphics[width=1\linewidth]{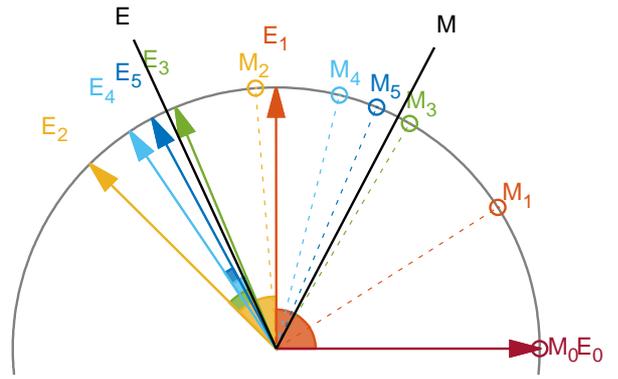}
\par\end{centering}
\caption{\label{fig:CORDIC_path}Example of five CORDIC rotations given $M=2-\sin2\approx62.49^{\circ}$
and $e=1$. Starting with $M_{0}=0$ and $E_{0}=0$, $M_{n}$ and
$E_{n}$ approach $M$ and $E=2\approx114.59^{\circ}$. The arcs indicates
the current rotation angle $\alpha_{n}$ and direction at each step.}
\end{figure}

\begin{figure}
\begin{centering}
\includegraphics[width=1\linewidth]{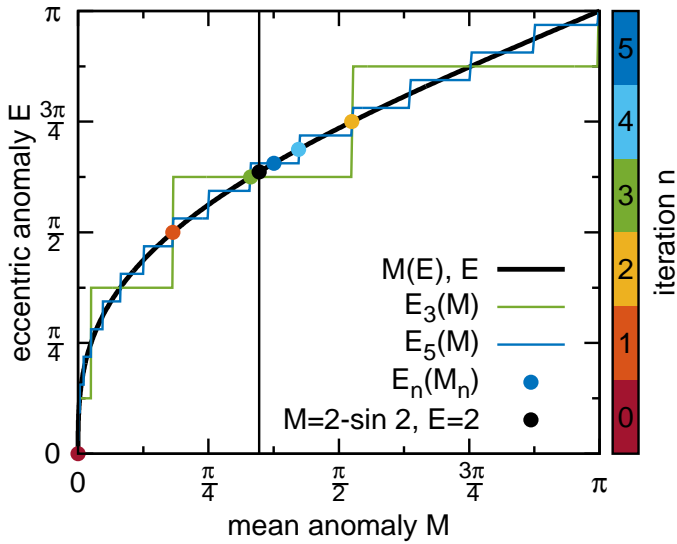}
\par\end{centering}
\caption{\label{fig:CORDIC}CORDIC-like approximation of Kepler's equation
($e=1$). The curves $E_{3}$ and $E_{5}$ illustrate the convergence
towards Kepler's equation after three and five iterations. The coloured points
$M_{n},E_{n}$ are passed during the iterations to find $E(M=2-\sin2)$
(same colour-coding as in Fig.~\ref{fig:CORDIC_path}).}
\end{figure}

Since the iteration starts with $E_{0}=0$, we simply have $c_{0}=\cos E_{0}=1$
and $s_{0}=\sin E_{0}=0$. The $\cos\alpha_{n}$ and $\sin\alpha_{n}$
terms can also be pre-computed using $\alpha_{n}$ from Eq.~(\ref{eq:alpha_n})
and stored in a table listing the triples $(\alpha_{n},\cos\alpha_{n},\sin\alpha_{n})$.
Now we have eliminated all function calls for sine and cosine of $E$
and $\alpha$ in the iteration. This is also true and important for
the $\sin E_{n}$ term in Eq.~(\ref{eq:sgn}) which is simply gathered
during the iteration process as $s_{n}$ via Eq.~(\ref{eq:rotE}).
Figures~\ref{fig:CORDIC_path} and \ref{fig:CORDIC} illustrate the convergence for input $M=2-\sin2\approx1.0907$ and $e=1$ towards
$E=2$.

In Appendix \ref{sec:python-code} we provide a short python code
which implements the algorithm described in this section. A slightly
more elegant formulation is possible with complex numbers $z_{n}=\exp(\imath E_{n})=\cos E_{n}+\imath\sin E_{n}$
and $a_{n}=\exp(\imath\alpha_{n})=\cos\alpha_{n}+\imath\sin\alpha_{n}$.
Then Eq.~(\ref{eq:rotE}) becomes $z_{n+1}=a_{n+1}z_{n}$ or $z_{n+1}=a_{n+1}^{*}z_{n}$
depending on $\sigma_{n}$ and in Eq.~(\ref{eq:sgn}) the substitution
$\sin E_{n}=\operatorname{\mathbb{I}m}(z_{n})$ has to be done.

A more detailed comparison with CORDIC is given Appendix~\ref{sec:CORDIC-sin}.
Before analysing the performance in Sect.~\ref{subsec:Performance},
we outline some optimisation possibilities in Sect.~\ref{sec:Variants-and-refinements}.

\section{\label{sec:Variants-and-refinements}Variants and refinements}

\subsection{\label{subsec:min-base}Minimal rotation base}

The algorithms in Sect.~\ref{sec:CORDIC-KE} and Appendix \ref{sec:CORDIC-sin}
use always positive or negative rotation directions. While after $n$
rotations the maximum error should be smaller then $\alpha_{n}$,
previous iterations can still be closer to the convergence limit.
In Fig.~\ref{fig:CORDIC_path}, for instance, $E_{3}$ is closer
to $E$ than $E_{4}$ and $E_{5}$. Consecutive rotations will compensate
this. A very extreme case is $M=0$, where $E_{0}=0$ is already the
exact value, but the first rotation brings it to $E_{1}=90^{\circ}$
followed by only negative rotation towards zero.

The CORDIC algorithm requires positive and negative rotations to have always the
same scale $K_{n}$ after $n$ iterations (Appendix \ref{sec:CORDIC-sin}).
But since we could not pull out $K_{n}$ from the iteration, we can
depart from this concept and allow for only zero and positive rotations
which could be called unidirectional \citep{2013Jain} or one-sided
\citep{Maharatna2004}. The decision to rotate or not to rotate becomes
\begin{equation}
\sigma_{n+1}=\begin{cases}
1 & (E_{n}+\alpha_{n+1})-e\sin(E_{n}+\alpha_{n+1})<M\\
0 & \text{else}.
\end{cases}\label{eq:sgn-01}
\end{equation}
Thereby we compose $E_{n}$ in Eq.~(\ref{eq:En}) by minimal
rotations and can expect also a better propagation of precision. Moreover,
while the term $\sin(E_{n}+\alpha_{n+1})=s_{n}c_{a}+c_{n}s_{a}$ still
needs to be computed in each iteration to derive for $\sigma_{n+1}$,
the term $c_{n+1}$ is updated only when $\sigma_{n+1}=1$ which will
be on average in about 50\% of the cases.
We note also that with the minimal rotation base the curve in Eq.~(\ref{eq:KE})
is approached from below.

\subsection{\label{subsec:CORDIC-Newton}CORDIC-like Newton method}

The convergence of the CORDIC-like algorithms in Sect.~\ref{sec:CORDIC-KE}
and \ref{subsec:min-base} is linear, while Newton's method is quadratic.
Of course, our algorithm can serve start values (and their cosine
and sine terms) for other root finders at any stage. The quadratic
convergence should bring the accuracy from, for example, $10^{-8}$ down
to $10^{-16}$ at the cost of mainly one division. We think this should
be preferred over the other possible optimisations mentioned in Appendix~\ref{sec:CORDIC-sin}.

Still, we do not want to lose the ability to provide the cosine and
sine terms after one (or more) Newton iterations. We can propagate
them simultaneously in a similar manner as in Eqs.~(\ref{eq:En+1})
and (\ref{eq:rotE}) but now using small angle approximations. Newton's
method (Eq.~(\ref{eq:Newton})) directly proposes a rotation angle
\begin{equation}
\alpha_{n+1}=E_{n+1}-E_{n}=\frac{M-E_{n}+es_{n}}{1-ec_{n}}.
\end{equation}
This means that there is no need to check for a rotation direction $\sigma_{n}$.
For small angles, there are the approximations $\sin\alpha\approx\alpha$
and $\cos\alpha\approx1$. If one works with double precision ($2^{-53}$),
the error $\epsilon(\alpha)=|1-\cos\alpha|<\frac{\alpha^{2}}{2}$
is negligible, meaning smaller than $2^{-54}\approx5.5\times10^{-17}$,
when $|\alpha|<7.5\times10^{-9}$ (given for $a_{29}$). Then we can
write
\begin{align}
c_{n+1} & =c_{n}+\alpha_{n+1}s_{n}\\
s_{n+1} & =s_{n}+\alpha_{n+1}c_{n}.
\end{align}

\subsection{\label{subsec:CORDIC-Halley}Halley's method}

Halley's method has a cubic convergence. Ignoring third order terms,
one can apply the approximations $\sin\alpha\approx\alpha$ and $\cos\alpha\approx1-\frac{\alpha^{2}}{2}$.
The error $\epsilon(\alpha)=|\alpha-\sin\alpha|<|\frac{\alpha^{3}}{6}|$
is negligible in double precision, when $\alpha<\sqrt[3]{6\times2^{-54}}=6.93\times10^{-6}$
given for $\alpha_{19}\approx5.99\times10^{-6}$. Similar to Sect.~\ref{subsec:CORDIC-Newton},
the iteration scheme with co-rotation for Halley's method is

\begin{align}
\alpha_{n+1} & =\frac{(1-ec_{n})(M-M_{n})}{(1-ec_{n})^{2}+\frac{1}{2}es_{n}(M-M_{n})}\\
s_{n+1} & = \left( 1-\frac{\alpha_{n+1}^{2}}{2} \right) s_{n}+\alpha_{n+1}c_{n}\\
c_{n+1} & = \left( 1-\frac{\alpha_{n+1}^{2}}{2} \right) c_{n}+\alpha_{n+1}s_{n}.
\end{align}

\section{\label{sec:CORDIC-KEh}Hyperbolic mode}

For eccentricities $e\ge1$, Kepler's equation is
\begin{equation}
M=e\sinh H-H.\label{eq:KEh}
\end{equation}
This case can be treated as being similar to the elliptic case, when the trigonometric
terms are replaced by hyperbolic analogues. Equations (\ref{eq:sgn}),
(\ref{eq:alpha_n}), and (\ref{eq:rotE}) become
\begin{align}
\sigma_{n+1} & =\begin{cases}
-1 & e\sinh H_{n}-H_{n}>M\\
+1 & \text{else}
\end{cases}\\
\alpha_{n} & =\frac{4\ln2}{2^{n}}\label{eq:alphah_n}\\
\left(\begin{array}{c}
c_{n+1}\\
s_{n+1}
\end{array}\right) & =\left[\begin{array}{cc}
\cosh\alpha_{n+1} & \sigma_{n+1}\sinh\alpha_{n+1}\\
\sigma_{n+1}\sinh\alpha_{n+1} & \cosh\alpha_{n+1}
\end{array}\right]\left(\begin{array}{c}
c_{n}\\
s_{n}
\end{array}\right).\label{eq:rotH}
\end{align}
where a sign change in Eq.~(\ref{eq:rotH}) leads to hyperbolic rotations.
This equation set covers a range of $H\in[-4\ln2,4\ln2]$. However,
$H$ extends to infinity. In the elliptic case, this is handled by
an argument reduction which makes use of the periodicity. In the hyperbolic
case, the corresponding nice base points $H_{0}$ are
\begin{align}
H_{0} & =m\ln2\label{eq:H0}\\
c_{0} & =\cosh H_{0}=\frac{1}{2}[\exp H_{0}+\exp(-H_{0})]=2^{m-1}+2^{-m-1}\\
s_{0} & =\sinh H_{0}=\frac{1}{2}[\exp H_{0}-\exp(-H_{0})]=2^{m-1}-2^{-m-1}.
\end{align}
For $m=0$, this becomes $H_{0}=0$, $c_{0}=1$, and $s_{0}=0$, in other words
similar to the elliptic case. For $m\ne0$, the start triple is
still simple to compute using only additions and bitshifts.

\begin{figure}
\includegraphics[width=1\linewidth]{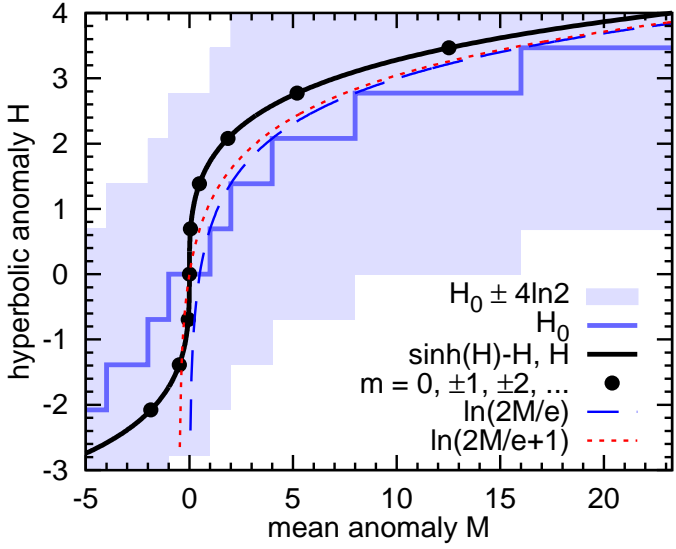}

\caption{\label{fig:Range-extension}Range extension for hyperbolic Kepler's
equation (black curve) in case $e=1$. The start value $H_{0}$ (blue
thick line) and the convergence domain of $\pm4\ln2$ (shaded area)
covers Kepler's equation. Nice base points (black dots) and two approximations
(dotted and dashed curves) are also indicated.}
\end{figure}

The main challenge is to obtain the integer $m$ from mean anomaly
$M$. In elliptic case, this needs a division with $2\pi$; in hyperbolic
case, it becomes a logarithmic operation. Figure~\ref{fig:Range-extension}
and Appendix~\ref{sec:Range-extension} show that
\begin{equation}
m=\sign M\cdot\max\left[0,\floor\left(1+\log_{2}\left|\frac{M}{e}\right|\right)\right]\label{eq:m}
\end{equation}
provides start values with the required coarse accuracy of better
than $4\ln2$. In floating point representation (IEEE 754), the second
argument in the maximum function of Eq.~(\ref{eq:m}) extracts simply
the exponent of $M/e$. In fixed point representation, it is the most
significant non-zero bit \citep{Walther:1971:UAE:1478786.1478840}.
This means a logarithm call is not necessary (cf. Appendix \ref{subsec:Hyperbolic-code}).

The hyperbolic iterations do not return $\cos H$ and $\sin H$, but
$\cosh H$ and $\sinh H$ which are the quantities needed to compute
distance, velocity, and Cartesian position.

\section{Accuracy and performance study}

\subsection{\label{subsec:Accuracy}Accuracy of the minimal rotation base algorithm}

The CORDIC-like algorithms have a linear convergence rate. The maximum
error for $E_{n}$ in iteration $n$ is given by $\alpha_{n}$. For
instance, we expect from Eq.~(\ref{eq:alpha_n}) $\alpha_{22}<10^{-6}$,
$\alpha_{29}<10^{-8}$ (single precision), $\alpha_{42}<10^{-12}$
and $\alpha_{55}<10^{-16}$ (double precision). To verify if double
precision can be achieved in practice, we forward-calculated with
Eq.~(\ref{eq:KE}) 1\,000 $(M(E),E)$ pairs uniformly sampled in
$E$ over $[0,\pi]$. Here $E$ might be seen as the true value. Then
we injected $M$ into our algorithms to solve the inverse problem
$E(M)$.

\begin{figure}
\begin{centering}
\includegraphics[width=1\linewidth]{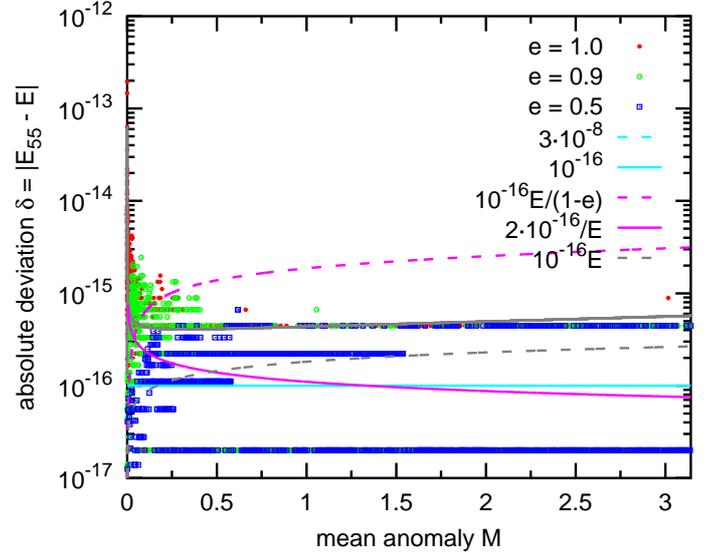}
\par\end{centering}
\caption{\label{fig:accuracy}Accuracy. For visibility, zero deviations ($\delta=0$)
within machine precision were lifted to $\delta=2\times10^{-17}$.}
\end{figure}

\begin{figure}
\begin{centering}
\includegraphics[width=1\linewidth]{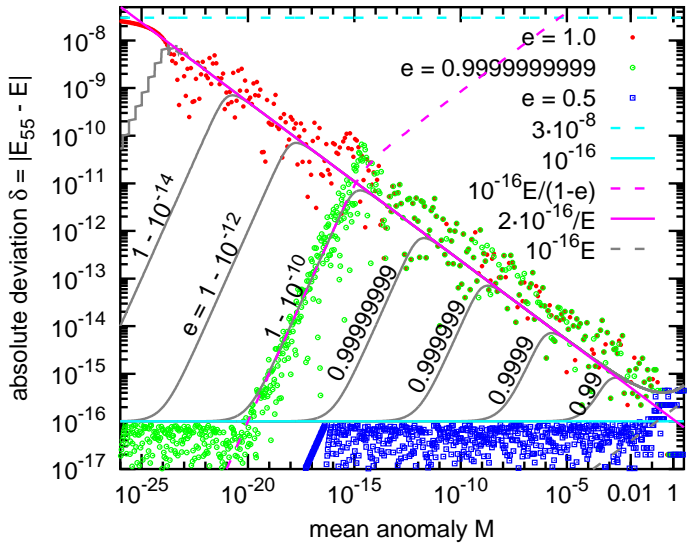}
\par\end{centering}
\caption{\label{fig:accuracy-logM}Similar to Fig.~4, but with logarithmic
scaling.}
\end{figure}

The differences between re-computed and true eccentric anomaly, $\delta_{55}=|E_{55}-E|$,
are shown in Fig.~\ref{fig:accuracy} for the cases $e=0.5$, 0.9,
and 1. For $M\gtrsim0.25$ and $e\le1$, the deviations are smaller
than $10^{-15}$ and close to machine precision $\epsilon$ as indicated
by the grey lines. Towards $M=0$ and $e>0.9$, the scatter increases.
Figure~\ref{fig:accuracy-logM} greatly magnifies this corner by
plotting the deviations against the logarithm of $E$ for $e=0.5$,
0.999\,999\,999\,9, and $1$.

The deviations $\delta_{55}$ become large in this corner, because
the inversion of Eq.~(\ref{eq:KE}) is ill-posed. For $e=1$, third
order expansion of Eq.~(\ref{eq:KE}) is $M=E-(E-\frac{1}{6}E^{3})=\frac{1}{6}E^{3}$.
If the quadratic term in $1-\frac{1}{6}E^{2}$ is below machine precision
$\epsilon$, it cannot be propagated numerically in the subtraction
and leads to a maximum deviation of $E<\sqrt{6\epsilon}\approx10^{-8}$
at $E=10^{-8}$ ($M=10^{-24}$).

\begin{figure}
\begin{centering}
\includegraphics[width=1\linewidth]{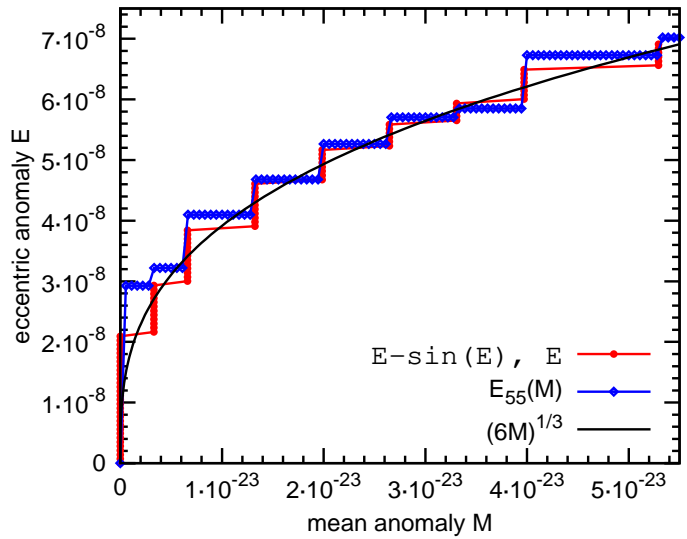}
\par\end{centering}
\caption{\label{fig:corner-root}The ill-posed Kepler equation in the corner
$M=0$ and $e=1$. Comparison of the computer approximations $M(E)=E-e\sin E$
(red) and our algorithm $E_{55}(M)$ (blue) with a cubic root (black;
from well-posed inversion of a third-order Taylor expansion).}
\end{figure}

Figure~\ref{fig:corner-root} illustrates the ill-posed problem.
For $e=1$ and very small $M$, the cubic approximation provides a
very accurate reference. For comparison, we overplot our $E_{55}$
solution as well as computer generated pairs $\texttt{E-sin(E),E}$.
Both exhibit a discretisation of $10^{-8}$ in $E$. We conclude that
Newton's, Halley's, and other methods which directly try to solve
Eq.~(\ref{eq:KE}) are as well only accurate to $10^{-8}$ in this
corner when operating with double precision! Therefore, the $M,E$
pairs mentioned above were not generated with built-in \texttt{sin()}-function
but using a Taylor expansion of Kepler's equations

\begin{equation}
M=E(1-e)-e\left(-\frac{E^{3}}{3!}+\frac{E^{5}}{5!}-\ldots\right).\label{eq:KE_Taylor}
\end{equation}
We may call the term in brackets rest function of sine. It differs by a factor of $-E^{-3}$ compared to its first naming in \citet{1959himm.book.....S}; see also Eq.~(\ref{eq:UKE_Taylor}).

We find that the deviations approximately follow the form
\begin{equation}
\delta_{55}(E,e)=|E_{55}-E|\approx10^{-16}+10^{-16}\frac{E}{1-e\cos E}+10^{-16}E\label{eq:Accuracy}
,\end{equation}
as indicated by the curves in Figs.~\ref{fig:accuracy} and \ref{fig:accuracy-logM}.
For very small $M$ and $E$, the constant term $10^{-16}$ dominates.
Then the middle term, which is related to the derivative of Eq.~(\ref{eq:KE}),
sets in. \mbox{Finally}, the linear term $10^{-16}E$ takes over which describes
the accuracy to which $E$ can be stored. The middle term leads to
an ascending and a descending branch which can be approximated by
$10^{-16}\frac{E}{1-e}$ and $10^{-16}\frac{2}{E}$ and forms a local
maximum. The two lines intersect at $E\approx\sqrt{2(1-e)}$ and $\delta_{55}(e)\approx10^{-16}\sqrt{\frac{2}{1-e}}$.
For highest resolvable eccentricity below one, this provides $\delta(e=1-10^{-16})\approx\sqrt{2}\times10^{-8}$.

\subsection{Accuracy of algorithm variants}

We find that the algorithm with positive and negative rotations ($E_{n}^{+-}$,
Sect.~\ref{sec:CORDIC-KE}) provides slightly worse accuracy for all
eccentricities (Fig.~\ref{fig:accuracy_of_algorithms}). For instance
for $e=0.9$, it is limited to $10^{-15}$ and it ranges up to $10^{-5}$
for $e=1$ at $M=10^{-16}$ ($E=10^{-5}$).

The combination of one final Newton iteration with $E_{29}$ as suggested
in Sect.~\ref{subsec:CORDIC-Newton} has a similar accuracy as $E_{55}$
(in particular at $e=1$) and provides therefore an alternative shortcut.
The combination of one Halley iteration with $E_{19}$ has
generally similar performance, but for high eccentricities it approaches $10^{-6}$.

\subsection{\label{subsec:Performance}Performance}

Before we discuss some performance results, we should mention that
a comparison of the algorithm speed is not simple. Depending on the
hardware, multiplications might be little or much more expensive than
additions. Then there is the question about the level of optimisation
done in the code, by the compiler, or in built-in routines.

It can be also instructive to count the number of floating point operations.
In Appendix~\ref{sec:python-code} we see five multiplications, four additions,
and one comparison in each iteration. For single precision this must be executed 29 times.
On the other hand, a Newton iteration in Eq.~(\ref{eq:Newton})
has one division, two multiplications, four additions, and one sine
and cosine term. In case the latter two terms are computed via Taylor
expansion to 17th order and with a Horner scheme, each will contribute
additional nine multiplications and nine additions. Furthermore, the cosine
and sine routines have to check the need for range reduction each
time. Actually, also an additional final sine and cosine operation
should be loaded to Newton's methods, since $\sin E$ and $\cos E$
are usually needed in subsequent calculations.

Given same weight to all listed operations, we count approximately ten operations
per CORDIC iteration vs. about per 43 operation per Newton iteration.
Then we estimate that four CORDIC iteration are as expensive as one
Newton iteration.

We have implemented both CORDIC-like and Newton's method in a C program.
The Newton's method used the start guess $E_{0}=M+0.85e$ and called
cos and sin function from the standard \texttt{math.h} library. The
run time was measured with python \texttt{timeit} feature for $\alpha_{29}$
($10^{-8}$) and for 1000 mean anomalies uniformly sampled from $0$
to $\pi$. The run-time of the CORDIC-like algorithm is independent
of $e$, while Newton's method is fast for small eccentricities and
slower for high eccentricities. We measured that our CORDIC-like method
has double, same, and half speed for $e=1$, 0.01, and 0, respectively,
or that 14 CORDIC-like iteration are as expensive as one Newton iteration.

\subsection{Comparison with \citet{1997CeMDA..66..309F}}

There are some similarities between this and the work of \citet{1997CeMDA..66..309F},
and a comparison can be enlightening and useful to put our method
in context. \citet{1997CeMDA..66..309F} uses a $128$ long table
sampled uniformly in $E$ which provides an accuracy of $1/128$.
In contrast, we need a table of length $\log_{2}128=7$ to obtain
this accuracy and higher accuracy is not limited by a huge table.
While \citet{1997CeMDA..66..309F} likely needs fewer iterations
due to the use of discretised Newton method, we have no divisions. 

\citet{1997CeMDA..66..309F} developed an iterative scheme with truncated
Taylor series of Newton's method to avoid recomputation of sine and
cosine angles. Similar, our methods outlined in Sects.~\ref{subsec:CORDIC-Newton}
and \ref{subsec:CORDIC-Halley} avoids such recomputations by using
truncated Taylor series for sine and cosine for small angles $\alpha_{n}$.
Our method appears less complex since it needs only three short equations.

\subsection{Universal Kepler's equation}

Our algorithms can solve eccentric and hyperbolic Kepler's equations.
The two are very similar and could be unified, similar to the method in \citet{Walther:1971:UAE:1478786.1478840}.
Both algorithm includes also the eccentricity $e=1$ (which are the
limits for rectilinear ellipse and hyperbola,
i.e. radial cases). The question is, therefore, whether our algorithm
can also be applied to universal Kepler's equation.

Following \citet{1999CeMDA..75..201F} and using the Stumpff function
of third degree, $c_{3}$ \citep{1959himm.book.....S}, the universal
Kepler's equation can be written as
\begin{align}
L & =G+\kappa G^{3}c_{3}(\lambda G^{2})\label{eq:UKE}\\
 & =G+\kappa\frac{\sqrt{\lambda}G-\sin\sqrt{\lambda}G}{\sqrt{\lambda}^{3}}\label{eq:UKE_trig}\\
 & =G+\kappa\left[\frac{G^{3}}{3!}-\frac{\lambda G^{5}}{5!}+\frac{\lambda^{2}G^{7}}{7!}-\ldots\right]\label{eq:UKE_Taylor}
\end{align}
with universal mean anomaly $L$, universal eccentric anomaly $G$,
Brunnow's parameter $\lambda=\frac{1-e}{1+e}$, and $\kappa=\frac{e}{1+e}$.
In parabolic case ($e=1$, $\lambda=0$), it becomes a cubic equation.
Using Eq.~(\ref{eq:UKE_trig}) and the relations $L=\frac{1+\lambda}{2\sqrt{\lambda}^{3}}M$
and $E=\sqrt{\lambda}G$, one can derive Kepler's equation, Eq.~(\ref{eq:KE})
\citep[see Appendix A in][]{1999CeMDA..75..201F}.

Equation~(\ref{eq:UKE_trig}) is expressed with a sine term. However,
the application of a CORDIC-like algorithm to solve for $G$ seems
not possible due to the scaling term $\sqrt{\lambda}$ inside the
sine function.

\section{Summary}

We show that the eccentric anomaly $E$ from Kepler's equation
can be computed by a CORDIC-like approach and is an interesting alternative
to other existing tools. The start vector $E_{0}=0^{\circ}$ and its
Cartesian representation ($\cos E_{0}=1$ and $\sin E_{0}=0$) are
rotated using a set of basis angles. The basis angles $\alpha_{n}$
and its trigonometric function values can be pre-computed and stored
in an auxiliary table. Since the table is short and independent of
$e$, it can be also hard-coded.

Our method provides $E$, $\sin E$, and $\cos E$ without calling
any transcendental function at run-time. The precision is adjustable
via the number of iterations. For instance, single precision is obtained
with $n=29$ iterations. Using double precision arithmetic, we found
the accuracy is limited to $\sim10^{-15}\sqrt{\frac{2}{1-e}}$ in
the extreme corner ($M=0$, $e=1$). For accuracy and speed
we re\-commend the one-sided algorithm described in Sect.~\ref{subsec:min-base}.

Our method is very flexible. As a stand-alone method it can provide
high accuracy, but it can be also serve start value for other refinement
routines and coupled with Newton's and Halley's method. In this context
we proposed in Sects.~\ref{subsec:CORDIC-Newton} and \ref{subsec:CORDIC-Halley}
to propagate cosine and sine terms simultaneously using small angle
approximations in the trigonometric addition theorems and derived
the limits when they can be applied without accuracy loss.

Though the number of iterations appears relatively large, the computational
load per iteration is small. Indeed, a with simple software implementation
we find a performance that is good competition for Newton's method.
However, CORDIC algorithms utilise their full potential when implemented
in hardware, that is, directly as a digital circuit. So-called field programmable
gate arrays (FGPA) might be a possibility to install our algorithm
closer to machine layout. Indeed, hardware oriented approaches can
be very successful. This was shown by the GRAvity PipelinE (GRAPE)
project \citep{1999Sci...283..501H}, which tackled N-body problems.
By implementing Newtonian pair-wise forces efficiently in hardware,
it demonstrated a huge performance boost and solved new astrodynamical
problems \citep{2003IAUS..208....1S}.

Though we could not completely transfer the original CORDIC algorithm
to Kepler's equation, it might benefit from ongoing developments and
improvement of CORDIC algorithms which is still an active field. In
the light of CORDIC, solving Kepler's equations appears almost as
simple as computing a sine function.

\begin{acknowledgements}
The author thanks the referee Kleomenis Tsiganis for useful comments
and M. K\"urster and S. Reffert for manuscript reading. This work
is supported by the Deutsche Forschungsgemeinschaft under DFG RE 1664/12-1
and Research Unit FOR2544 "Blue Planets around Red
Stars", project no. RE 1664/14-1.
\end{acknowledgements}

\bibliographystyle{aa}
\bibliography{ke_cordic}

\appendix

\section{\label{sec:python-code}Illustrative python code}

The following codes implement the two-sided algorithms described in
Sects.~\ref{sec:CORDIC-KE} and \ref{sec:CORDIC-KEh}.

\subsection{Elliptic case}

\noindent 

\noindent \definecolor{blue}{rgb}{0,0,1.0}
\definecolor{deepblue}{rgb}{0,0,0.5}
\definecolor{deepred}{rgb}{0.6,0,0}
\definecolor{deepgreen}{rgb}{0,0.4,0}
\definecolor{coldigit}{RGB}{176,128,0} 
\definecolor{grey}{RGB}{136,135,134} 

\lstset{
language=python,
basicstyle=\ttfamily,
numbers=left,
numberstyle=\tiny,
numbersep=5pt,
 linewidth=\linewidth,
xleftmargin={0.05\linewidth},
commentstyle=\color{grey},
frame=tb,
keywordstyle=\bfseries\color{deepgreen},
keywords=[2]{max,range}, keywordstyle=[2]\color{deepgreen},
emph={math},             emphstyle=\bfseries\color{blue},
emph=[2]{Ecs,Hcs},       emphstyle=[2]\color{blue},    
frame=tb,                         
alsoletter={0,1,2,3,4,5,6,7,8,9,.},
keywords=[5]{0,1,2,3,29,60,0.,1.,0.5},
keywordstyle=[5]\color{coldigit},
literate=%
   {,}{,}1
}

\begin{lstlisting}[linerange={9-15,33-41}]
# initialise cos-sin look-up table
from math import pi, cos, sin, floor
divtwopi = 1 / (2*pi)
acs = [(a, cos(a), sin(a)) for a in \
        [pi/2**i for i in range(1,60)]]

def Ecs(M, e, n=29):
   E = 2 * pi * floor(M*divtwopi+0.5)
   cosE, sinE = 1., 0.
   for a,cosa,sina in acs[:n]:
      if E-e*sinE > M:
         a, sina = -a, -sina
      E += a
      cosE, sinE = cosE*cosa - sinE*sina,\
                   cosE*sina + sinE*cosa
   return E, cosE, sinE
\end{lstlisting}

As an example, calling the function with \texttt{Ecs(2-sin(2),~1)}
should return the triple \texttt{(1.99999999538762, -0.4161468323531165,
0.9092974287451092)}.

We note that this pure python code illustrates the functionality and
low complexity (not performance). An implementation in C (including
a wrapper to python) and gnuplot with algorithm variants (e.g. one-sided
for improved accuracy and speed) are available online\footnote{\url{https://github.com/mzechmeister/ke}}.

\subsection{\label{subsec:Hyperbolic-code}Hyperbolic case}

\begin{lstlisting}[linerange={44-50,70-81}]
# initialise cosh-sinh look-up table
from math import log, cosh, sinh, frexp, ldexp
ln2 = log(2)
acsh = [(a, cosh(a), sinh(a)) for a in \
        [4*ln2/2**i for i in range(1,60)]]

def Hcs(M, e, n=29):
   m = max(0, frexp(M/e)[1])
   if M < 0: m = -m
   H = m * ln2
   coshH = ldexp(1, m-1) + ldexp(1, -m-1)
   sinhH = ldexp(1, m-1) - ldexp(1, -m-1)
   for a,cosha,sinha in acsh[:n]:
      if e*sinhH-H > M:
         a, sinha = -a, -sinha
      H += a
      coshH, sinhH = coshH*cosha + sinhH*sinha,\
                     coshH*sinha + sinhH*cosha
   return H, coshH, sinhH
\end{lstlisting}

Calling \texttt{Hcs(sinh(2)-2,~1)} returns the triple \texttt{(1.9999999991222275,
3.7621956879000753, 3.626860404544669)}.

The function \texttt{frexp} returns the mantissa and the exponent
of a floating point number. Vice versa, the function \texttt{ldexp}
needs a mantissa and an exponent to create a floating point number.

\section{\label{sec:CORDIC-sin}Comparison with original CORDIC algorithm
for the sine}

The CORDIC algorithm for the sine function is in practice further
optimised to work solely with additive operation and bitshift operation
and without multiplication or even division. The rotation angles $\alpha_{n}$
are slightly modified for this purpose such that now for $n>1$
\begin{equation}
\tan\alpha_{n}=\frac{1}{2^{n-2}}.\label{eq:tana}
\end{equation}

As in Sect.~\ref{sec:CORDIC-KE}, the first iteration
starts with $\alpha_{1}=90^{\circ}$ ($\tan\alpha_{1}=\infty$). The
second angle is also the same $\alpha_{2}=45^{\circ}$ ($\tan\alpha_{2}=1$),
while all other angles are somewhat larger than in Eq.~(\ref{eq:alpha_n})
and the convergence range is $\sum_{n=1}^{\infty}\alpha_{n}=1.0549\pi=189.9^{\circ}$.
Using Eq.~(\ref{eq:tana}) we can write Eq.~(\ref{eq:rotE}) as
\begin{equation}
\left(\begin{array}{c}
c_{n+1}\\
s_{n+1}
\end{array}\right)=\cos\alpha_{n+1}\left[\begin{array}{cc}
1 & -\sigma_{n+1}\frac{1}{2^{n-1}}\\
\sigma_{n+1}\frac{1}{2^{n-1}} & 1
\end{array}\right]\left(\begin{array}{c}
c_{n}\\
s_{n}
\end{array}\right).\label{eq:rotE_atan}
\end{equation}

The next move in CORDIC is to take out the factor $\cos\alpha_{n}$
from this equation and to accumulate it separately. This saves two
multiplications, but changes the magnitude of the vector by $\cos\alpha_{n}$
in each iteration and after $n$ iteration by
\begin{equation}
K_{n}=\prod_{i=2}^{n}\cos\alpha_{i}=\prod_{i=2}^{n}\frac{1}{\sqrt{1+\tan^{2}\alpha_{i}}}=\prod_{i=0}^{n-2}\frac{1}{\sqrt{1+4^{-i}}}~.
\end{equation}
This scale correction factor $K_{n}$ can be applied once after the
last iteration. The iteration scheme for the scale dependent vector
$X_{n},Y_{n}$ (where $c_{n}=K_{n}X_{n}$ and $s_{n}=K_{n}Y_{n}$)
becomes 
\begin{align}
X_{n+1} & =X_{n}-Y_{n}\frac{\sigma_{n+1}}{2^{n-1}}\\
Y_{n+1} & =Y_{n}+X_{n}\frac{\sigma_{n+1}}{2^{n-1}}.
\end{align}
This could be applied for the solution of Eq.~(\ref{eq:KE}), too.
However, the check for rotation direction in Eq.~(\ref{eq:sgn})
requires then a scale correction ($E_{n}-eK_{n}X_{n}>M$). So one
multiplication remains, while one multiplication can still be saved.
The additional multiplication with $e$ could be saved by including
it into the start vector (i.e. $z_{0}=e\exp(\imath E_{0})=e=X_{0}$
for $E_{0}=0$). 

The final move in CORDIC is to map float values to integer values
which has the great advantage that the division by a power of two can
replaced by bit shifts. But since we could not eliminate all multiplications,
the full concept of CORDIC is unfortunately not directly applicable
to solve Kepler's equation. Therefore, we call our method CORDIC-like.

However, it should be noted that also scale-free CORDIC algorithms
have been developed \citep{2016arXiv160602468H} at the expense of
some additional bitshift operations. Furthermore, the scale correction
$K_{n}$, which converges to $K_{\infty}\approx0.607253\approx1/1.64676$,
does not change in double precision for $n=26$. This means from iteration
$n=26$ onwards, the concept of CORDIC is applicable, when setting
the magnitude of the start vector to $K_{26}e$. Similar, for $n=29$,
the cosine term becomes one in double precision ($\cos\alpha_{n}=1$),
which also indicates a quasi-constant scale factor and means that
we can save this multiplication. In Sects.~\ref{subsec:CORDIC-Newton}
and \ref{subsec:CORDIC-Halley} we exploit this fact in another way.

\section{Accuracy plots}

\begin{figure*}
\includegraphics[width=0.5\linewidth]{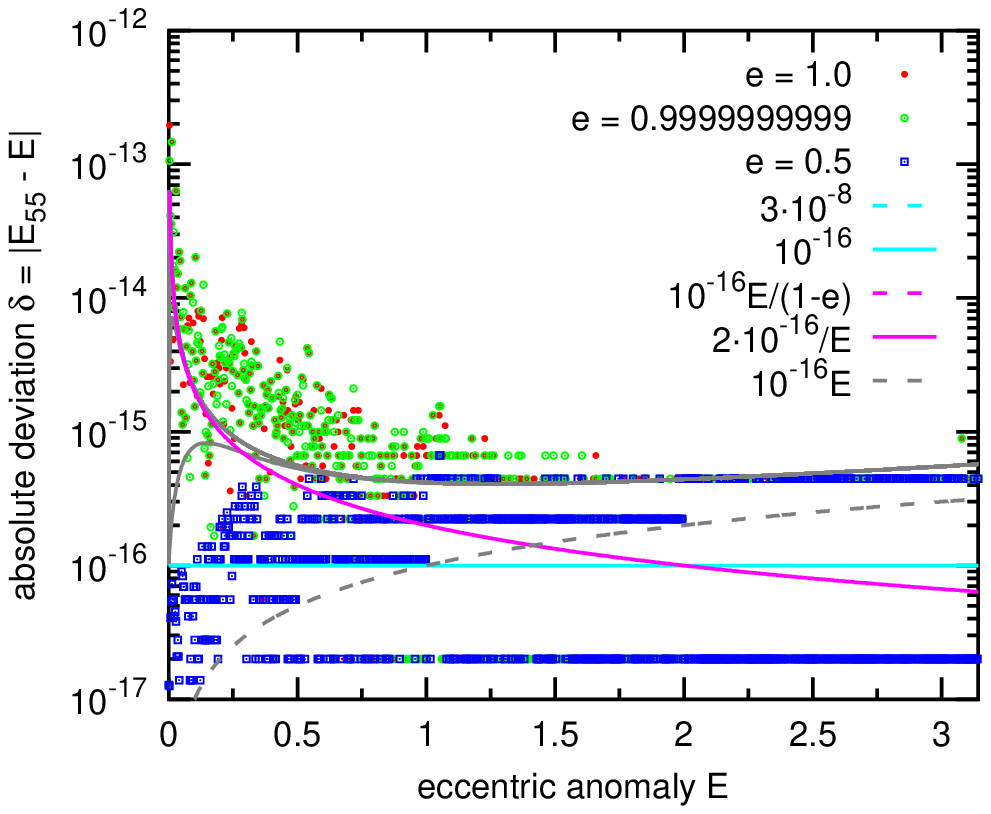}\includegraphics[width=0.5\linewidth]{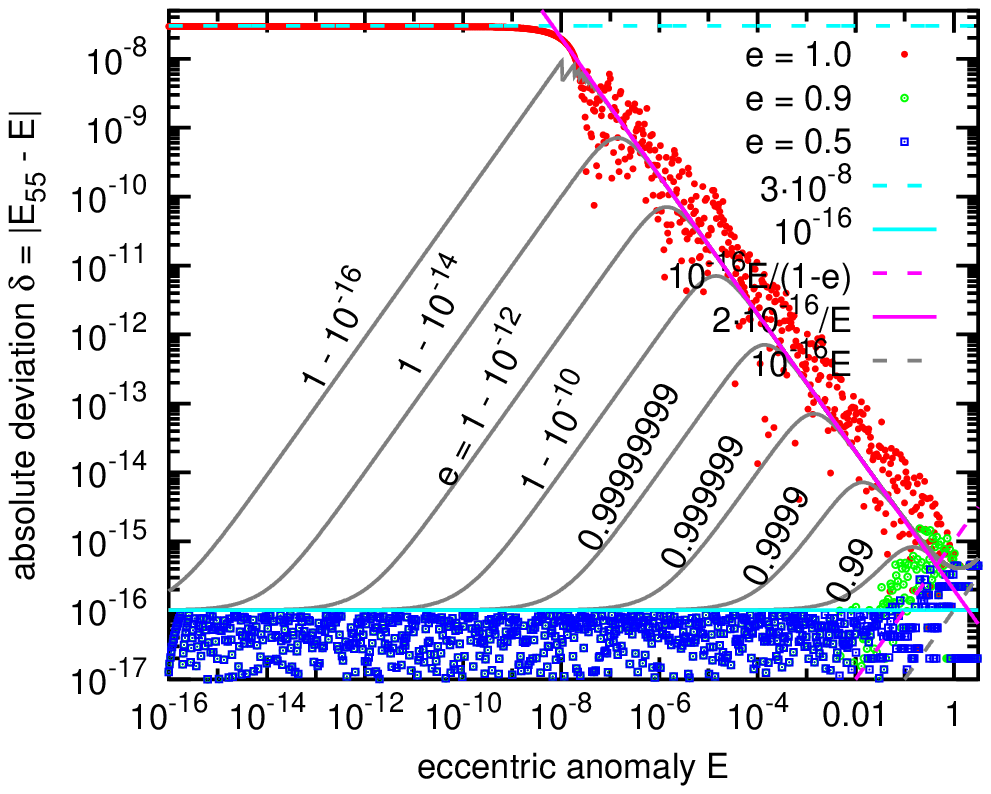}

\caption{Similar to Figs.~\ref{fig:accuracy} and \ref{fig:accuracy-logM},
but as function of $E$.}
\end{figure*}
\begin{figure*}
\includegraphics[width=0.5\linewidth]{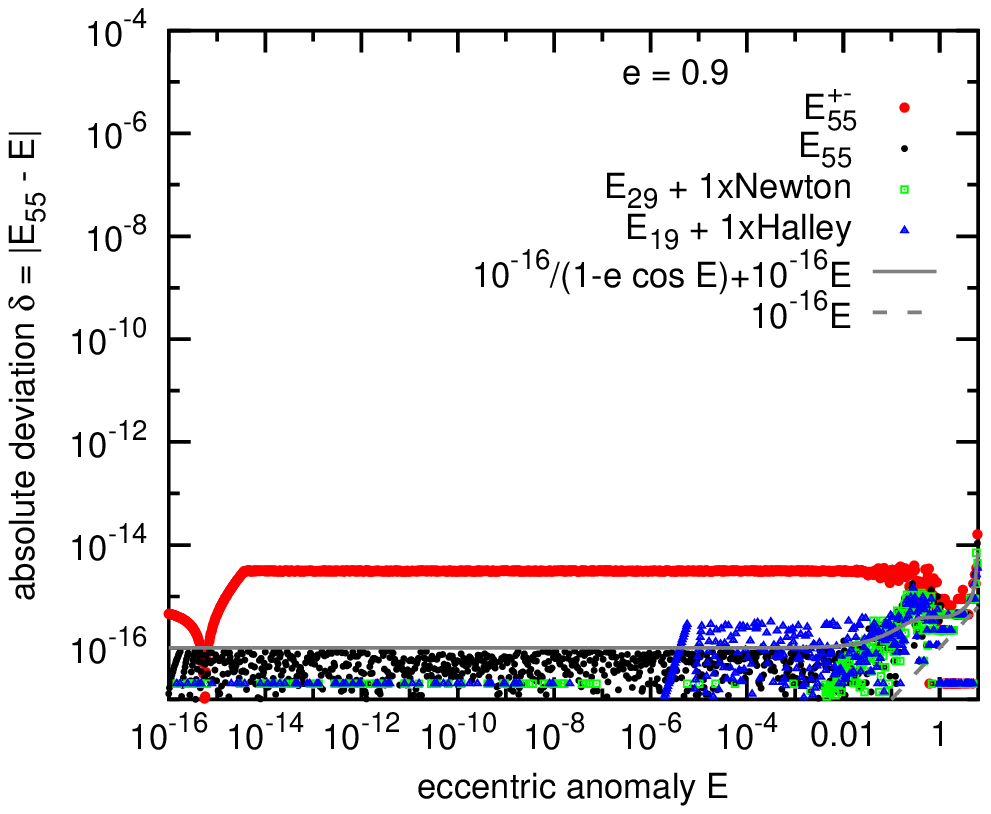}\includegraphics[width=0.5\linewidth]{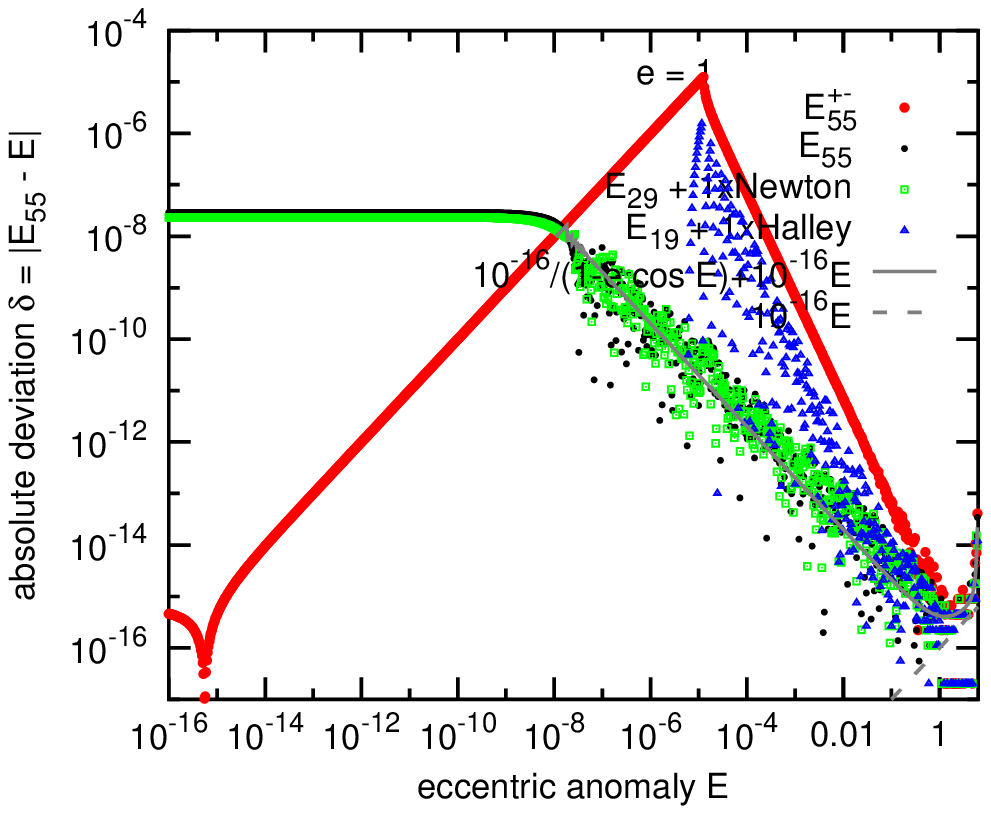}

\caption{\label{fig:accuracy_of_algorithms}Accuracy comparison of algorithm
variants.}
\end{figure*}

\section{\label{sec:Range-extension}Starting estimate for the hyperbolic case}

It can be shown that Eq.~(\ref{eq:m}) always provides a lower bound
for $H$ in case of $M\ge0$. Starting with Kepler's equation Eq.~(\ref{eq:KEh})
and using the identity $\sinh x=\frac{1}{2}[\exp x-\exp(-x)]$ and
the inequation $\exp(-x)\ge1-2x$ (for $x\ge0$), we obtain
\begin{align*}
M & =e\sinh H-H=\frac{e}{2}\left[\exp H-\exp(-H)-2\frac{H}{e}\right]\\
 & \le\frac{e}{2}\left[\exp H-1+2H\left(1-\frac{1}{e}\right)\right]\\
 & \le\frac{e}{2}[\exp H-1].
\end{align*}
For large values $H$, the approximation becomes more accurate. Further
reforming and introduction of the binary logarithm yields
\begin{align}
H & \ge\ln\left(\frac{2M}{e}+1\right)=\ln2\cdot\log_{2}\left(\frac{2M}{e}+1\right).\label{eq:Happrox_close}
\end{align}
The right side of Eq.~(\ref{eq:Happrox_close}) is plotted in Fig.~\ref{fig:Range-extension}
as a red dotted line. It belongs to the family of start guesses given
in \citet{1983CeMec..31..317B} ($H_{0}=\ln(\frac{2M}{e}+k)$). For
$\frac{M}{e}=\frac{1}{2}$, $e=1$, and $H_{0}=0$, the deviation
is $H-H_{0}\approx1.396=2.014\ln2$. Since, unfortunately, the pre-factor
is slightly larger than two, the convergence range is expanded to
$4\ln2$ by setting $\alpha_{1}=2\ln2$ in Eq.~(\ref{eq:alphah_n}).

For this large convergence range, we can further simply the start
guess by omitting the additive one in Eq.~(\ref{eq:Happrox_close})
\begin{align}
H & >\ln\frac{2M}{e}=\ln2\cdot\log_{2}\frac{2M}{e}=\ln2\cdot\left(1+\log_{2}\frac{M}{e}\right)\label{eq:Happrox_crude}\\
 & >\ln2\cdot\floor\left(1+\log_{2}\frac{M}{e}\right)\label{eq:Happrox_floor}
\end{align}
which finally leads to the start value for $m$ in Eq.~(\ref{eq:m}).
The right function in Eq.~(\ref{eq:Happrox_crude}) is plotted is
in Fig.~\ref{fig:Range-extension} as a blue dashed line. At $\frac{M}{e}=1$
and $e=1$, the hyperbolic anomaly is $H\approx1.729=2.495\ln2$ and
thus inside our convergence domain.

The simple start guess developed here is valid for all $e\ge1$ and
a lower bound for $M\ge0$, meaning that it is suitable for one-sided algorithms.

\end{document}